# Electronic structure and phase transition in polar $ScFeO_3$ from First Principles Calculations

Bog G. Kim[1,2*], M. Toyoda[3], Janghee Park[1], and Tamio Oguchi[2,4]

[1]Department of Physics, Pusan National University, Pusan, 609-735, South Korea
[2]Institute of Scientific and Industrial Research, Osaka University, Ibaraki, Osaka 567-0047, Japan
[3]Department of Physics, Tokyo Institute of Technology, Meguro-ku, Tokyo 152-8550, Japan
[4]CREST-JST, Kawaguchi, Saitama 332-0012, Japan

The properties of newly discovered polar $ScFeO_3$ with magnetic ordering are examined using *Ab initio* calculations and symmetry mode analysis. The GGA+*U* calculation confirms the stability of polar *R*3*c* Phase in $ScFeO_3$ and the pressure induced phase transition to non-polar *Pnma* phase. Octahedron tilting and structural properties as a function of applied pressure have been analyzed. The origin of polar phase is associated with instability of non-polar $R\bar{3}c$ phase and group theory using the symmetry mode analysis has been applied to understand this instability as well as the spontaneous polarization of polar *R*3*c* phase. The magnetic phase transition shows G-type AFM ordering of $Fe^{3+}$ ion within Goodenough-Kanamori theory and the possibility of magnetic spin structure has been analyzed by using energy analysis including spin canting possibility.

I. Introduction

Recently, there has been considerable interest in $ABO_3$-type perovskite compound, not only for practical applications, but also for their fundamental physics [1-9]. $ABO_3$ perovskite oxides with B-site transition metal have been studied quite a lot because of rich variety of physical properties such as ferroelectricity, superconductivity, and ferromagnetism [9-17]. In the structural point of view, $BO_6$ corner-sharing octahedra are linked three dimensionally, whereas A-site cation acts as building block for $BO_6$ octahedra [18]. One of the important empirical parameters for $ABO_3$ perovskite oxide is tolerance factor defined using Shannon ionic radii [19].

Ideal cubic perovskite is formed when the tolerance factor becomes unity. However, in the real situation, the deviation of tolerance factor will occur depending on constituent materials as well as on the external parameters such as pressure or temperature. Upon the deviation, the corner-sharing octahedra exhibit tilting in three dimensional patterns, which is categorized by Glazer notation [20-26]. The octahedral tilting can be expressed by using symbol of $a^{\#}b^{\#}c^{\#}$, in which the literals refer to tilt around the [100], [010], and [001] directions of the cubic perovskite and amplitude of tilting, and the superscript # takes the value 0, +, or – to indicate no tilt or tilts of successive octahedral in the same or opposite sense.

One of the most recent developments in the perovskite oxide is discovery of ferroelectricity and magnetic ordering in $ScFeO_3$ material [27, 28]. It is quite challenging to make $ScFeO_3$ materials because $Sc^{3+}$ ion is thought to be "too small" for A-site cation. When the material is synthesized under ambient conditions, random distribution of $Sc^{3+}$ and $Fe^{3+}$ is inevitable for A-site and B-site. Due to such randomness, the $ScFeO_3$ is known to be non-perovskite phase of bixbyite-type structure. M. –R. Li *et. al.*[27] reported that the polar corundum structure can be stabilized by combining high pressure and high temperature. They also showed the polar nature of corundum structure and calculated the polarization value of 3.3 uC/cm$^2$, which is quite small compared to that of BiFeO3 ( ~ 60 uC/cm$^2$) [29]. The magnetic measurement shows weak ferromagnetism, which is AFM ordering with small degree of spin canting. T. Kawamoto *et. al.* [28] questioned the "polar corundum phase" in the $ScFeO_3$ and have found the stability of orthorhombic phase using high pressure (~ 15 GPa) and high temperature (800 °C) by using in situ X-ray diffraction. They also showed that the orthorhombic phase (*Pnma*) of $ScFeO_3$ transformed into *R3c* phase with cooling. By using scanning transmission electron microscopy, they showed the fully ordered arrangement of $Sc^{3+}$ and $Fe^{3+}$ ions within the *R3c* phase. Analyzing neutron diffraction and susceptibility measurement, they also confirmed the weak ferromagnetism resulting from AFM ordering of $Fe^{3+}$ ion with small degree of spin canting.

In this study, we would like to report our results on the first principles calculation of $ScFeO_3$. The crystal structures of $ScFeO_3$ have been explored in the orthorhombic phase as well as the hexagonal phase. We confirmed the stability of *R3c* phase by density functional theory (DFT) with on-site *U* addition for Fe ion. The pressure induced phase transition from *R3c* phase to *Pnma* orthorhombic phase is also confirmed from our DFT calculations. We also analyzed the detailed structural change as a function of applied pressure. We also have studied the stability of *R3c* phase over non-polar *R$\bar{3}$c* phase. The symmetry mode group theory is used to explain the primary order parameter for such non-polar (ferroelectric) phase transition. Using the polarization calculation by berry phase method, we showed that the *R3c* phase has spontaneous polarization of ~100 uC/$cm^2$. Finally, the magnetic structure and spin canting within *R3c* phase are presented.

II. Calculation Details

We have performed the first-principles calculations with the Perdew-Burke-Ernzerhof generalized gradient approximation (PBE GGA) [30] with on-site Coulomb interaction (GGA+*U* method) [31] to the spin density functional theory and the projector augmented wave method as implemented in VASP [32, 33]. We considered the following valence electron configurations for atoms: $3s^23p^64s^03d^3$ for Sc, $3d^74s^1$ for Fe, and $2s^22p^4$ for Oxygen. In order to include correlation effect for transition metal Fe, we have applied on-site Coulomb interaction term of *U* with 4 eV [31, 34]. Electronic wave functions are expanded with plane waves up to a kinetic energy cutoff of 400 eV except for structural optimization, where 600 eV has been applied to reduce the effect of Pulay Stress. Momentum space integration is performed using $9 \times 9 \times 9$ Γ-centered Monkhorst-Pack k-point mesh [35] in two perovskite unit cell system or equivalent k density for larger supercell.

With the given symmetry of perovskite and antiferromagnetic (AFM) spin ordering imposed [27, 28], lattice constants as well as internal coordinates have been fully optimized until the residual Hellmann-Feyman forces became smaller than $10^{-3}$ eV/Å. It is interesting to note that insulating ground state is realized both with *U* ( = 4 eV) and with AFM spin configuration. The ISOTROPY and AMPLIMODES programs have been utilized to check the group-subgroup relationship and to quantify mode contribution to the total energy as well as the analysis of structural information [36, 37]. The spontaneous polarization of a fully distorted *R3c* phase is obtained by using berry phase method [38]. We also calculate the possibility of spin canting by using the spin non-collinear calculation. Since experimental data of spin canting (~ 0.2 °) is relatively small from antiferromagnetic (AFM) state [28], we start from AFM optimized structure to check the spin canting ground state. Our calculation is based on full Brillouine zone integration of spinor representation of Kohn-Sham Hamiltonian [39] implemented in VASP.

III. Results and Discussion

A. Structures and Equation of state

Figures 1(a), 1(b), and 1(c) depict the crystal structure of ScFeO3 in $R\bar{3}c$, $R3c$, and *Pnma* structure showing the tilting pattern. By assuming perfect 1:1 ordering of Sc and Fe atoms in A and B sites of perovskite structure, Fe and Oxygen form $FeO_6$ octahedra, which are linked three-dimensionally. The tilting pattern presented in ScFeO$_3$ can be described using the Glazer notation. In $R\bar{3}c$ and $R3c$ hexagonal structure, the tilting of octahedra can be described by $a^-a^-a^-$, while in *Pnma* orthorhombic structure, the $a^-a^-c^+$ tilting is realized. Note that $R3c$ structure can be considered as the distorted and shifted structure of a parent para-electric $R\bar{3}c$ structure (depicted in Fig. 1(d)).

The energetics of the phases is checked by the total energy vs. volume plot of ScFeO$_3$ by assuming G-type antiferromagnetism (AFM) and adding on-site Columnb term (*U*) for Fe ion of 4 eV (Fig. 1(e)). At each volume, full optimization of cell shape and lattice parameters, and internal coordinates of atoms have been performed with given symmetry. The lowest energy state of ScFeO$_3$ within AFM configuration is *R3c* hexagonal structure (space group number 160), which is same as LiNbO$_3$ [40] and LiOsO$_3$ compound [41] and the space group of previous experiment [27, 28]. The energy difference between *R3c* and *Pnma* structure is calculated to be 36.45 meV/f.u. and that of *R3c* and $R\bar{3}c$ is 308.1 meV/f.u. Both *R3c* phase and *Pnma* phase turn out to be stable in our preliminary phonon calculation (not shown). The structural parameters, such as lattice constants and internal coordinates of atoms are summarized in Table. 1. Our GGA+*U* calculations are in good agreement with experimental values of previous report [27, 28]. It is interesting to note that the *Pmna* structure can be stabilized by smaller volume, which means possible structural phase transition by applying pressure, which is correlated with high pressure synthesis of 1:1 ordered orthorhombic ScFeO$_3$ in the previous experiment [27, 28].

B. Pressure induced phase transition

In order to check the pressure-dependent phase transition, we have calculated the enthalpy of system as a function of applied pressure in Fig. 2(a). We have used isotropic pressure data of our first-principles calculation and again cell shape and internal coordinates of atoms are fully optimized within given pressure for *Pnma* and *R3c* structure. As can be seen from Fig. 2(a), the pressure dependence of enthalpy is almost linear function of the pressure. The *R3c* structure is stable at atmospheric pressure and it transformed to *Pnma* orthorhombic phase around 37.4 kbar. Therefore, experimental synthesis of the orthorhombic structure with 15 GPa (~ 150 kbar) pressure as well as low pressure stability of *R3c* phase can be understood from our first-principles calculation [27, 28].

With the given pressure, the structural parameters such as unit cell volume (per formula unit) and lattice constant are depicted in Fig. 2(b) and 2(c). By increasing pressure, the cell volume decreases monotonically by maintaining *R3c* structure (Fig. 2(b)) and, around 37.4 kbar, the abrupt volume change is observed, which correspond to the pressure induced phase transition to *Pnma* orthorhombic phase. We also have calculated pseudo-cubic lattice parameters of *R3c* and *Pnma*.

For *R3c* phase, the pseudo-cubic lattice parameter $a_0$ is related with lattice parameter of *R3c* phase divided by square root 2 ( $a_0 = a/\sqrt{2}$ ), whereas for *Pnma* phase, the pseudo-cubic lattice parameters, $a_0$, $b_0$, and $c_0$ are given by $a/\sqrt{2}$, $b/\sqrt{2}$, and $c/2$. As can be seen from Fig. 2(c), the lattice parameter of *R3c* phase decreases monotonically by applying external pressure and it abruptly decreases at phase transition pressure of 37.4 kbar. After phase transition, three lattice parameters representing orthorhombic structure can be observed and they decrease monotonically with pressure. It is interesting to note that the slope of $c_0$ lattice parameters is about two times larger than that of $a_0$ and $b_0$. The tilting angle of Glazer notation can be extracted from the lattice parameters and Wyckoff position of atoms and we have plotted $a^-$ tilting angle of *R3c* phase and $a^-$ and $c^+$ tilting angles of *Pnma* in Fig. 2(d). The tilting angle ($a^-$) decreases slightly but monotonically within *R3c* phase by applying pressure and it jumped up to two different values ($a^-$ and $c^+$) at critical pressure of phase transition. After the orthorhombic phase transition, the two tilting angles of $a^-$ and $c^+$ decrease monotonically with pressure.

## C. Mode analysis and polarization

We are now in a position to discuss the polar nature of *R3c* phase. In order to understand the nature of *R3c* phase, we have also done structural analysis of parent phase with *R$\bar{3}$c* structure with inversion symmetry. By having optimized structures in *R3c* and *R$\bar{3}$c* phase, the mode analysis has been performed with ISOTROPY and AMPLIMODES programs [36, 37]. Figure 3(a) and 3(b) depict two modes related with non-centrosymmetric phase transition. The $\Gamma_1^+$ mode is symmetry-preserving mode, which is related with rotation of octahedra perpendicular to the [111] direction. The phase transition to non-centrosymmetric *R3c* phase is associated with $\Gamma_2^-$ mode, which is mode of Fe-O bending with Sc displacement. The two modes are also important for the phase transition in polar $LiNbO_3$ and $LiOsO_3$ compounds [41].

The Figure 3(c) depicts the total energy vs. amplitude of modes and Table 2 summarized the amplitude of each modes as well as the atomic positional shift related with each modes. In fig. 3(c), the energy associated with coupled $\Gamma_1^+$ and $\Gamma_2^-$ modes is indexed as all modes (black circles) and the energies of individual $\Gamma_1^+$ and $\Gamma_2^-$ modes are plotted as green squares and blue diamonds, respectively. First of all, the energy of all modes is nearly same as that of $\Gamma_2^-$ mode, meaning that $\Gamma_2^-$ mode is primary order parameter of the phase transition. Secondly, the energy as a function of individual $\Gamma_1^+$ mode does not show any double-well nature, meaning the instability of phase within $\Gamma_1^+$ mode. However, after the system is displaced according to minimum energy state with $\Gamma_2^-$ mode and the energy associated with additional $\Gamma_1^+$ mode shows minimum point with finite amplitude, as shown in green squares in Fig. 3(c). As a consequence, the coupling of $\Gamma_2^-$ mode and $\Gamma_1^+$ mode is important to understand fictitious phase transition from *R$\bar{3}$c* to *R3c* phase. It is worth to note that the symmetry nature of phase transition is same as that observed in other perovskites such as $LiNbO_3$ and $LiOsO_3$ [41].

Spontaneous polarization is directly related with amplitude of $\Gamma_2^-$ mode, since $\Gamma_2^-$ mode is symmetry breaking transition. The value of spontaneous polarization has been obtained by using Berry phase method and depicted in Fig. 3(d). We confirm that the direction of spontaneous polarization is [111] direction as can be estimated from mode plot of Fig. 3(b). When we calculate the spontaneous polarization value, we also have utilized the standard conventional cell geometry to confirm the value. In the standard conventional cell geometry, the spontaneous polarization value is along the *c*-axis. The spontaneous polarization is multi-valued as shown in figure and the polarization quanta ($P_Q$) is about 68.2 μC/cm$^2$ for and the spontaneous polarization ($P_S$) is 100.4 μC/cm$^2$ [28]. It is worthwhile to note that the GGA+*U* (4 eV) and AFM configuration is used for this calculation with the optimized geometry. When we change the FM configuration of spin structure, the value of spontaneous polarization ($P_S$) is 99.8 μC/cm$^2$, indicating the small but non-zero magneto-electric effect can be realized in this compound.

### D. Magnetic and electronic structure

Now let us turn our attention to the magnetic phase of ScFeO$_3$. In Fig. 4(a), the equations of states with two different magnetic configurations are shown. At each given state, the lattice constant and internal coordinates of atoms are fully optimized starting from initial magnetic state and space group symmetry. We have used GGA+*U* (4 eV for Fe ion) for consistency. As can be clearly seen from this, the antiferromagnetic (AFM) configurations are always lower energy state compared to ferromagnetic (FM) states. The energy difference ($\Delta E$) between two magnetic states is 162.8 meV/f.u when the lowest states are compared. The magnetic transition temperature estimated from our calculation would be around 977 K ( $T_c \sim \Delta E/2$ ). It is interesting to note that the experimentally observed phase transition is around 500 K [27, 28]. This can be explained that the transition temperature of magnetic system is quite sensitive to the defects and any disordering of Sc and Fe might have critical effect on transition temperature as indicated in previous experimental reports [27, 28].

The nature of AFM structure can be understood from Goodenough-Kanamori super-exchange mechanism [42]. In the simple ionic picture, the Fe ion can be 3+ state with $d^5$ configuration. Each Fe ion is connected by sharing oxygen ion. Therefore we are dealing with exchange interaction of Fe $d^5$ – Oxygen – Fe $d^5$ system, which is quite well known situation where Goodenough-Kanamori super-exchange type AFM interaction can be applied. To explore this possibility in a real calculation, we have analyzed band structure and density of state within AFM configuration. Figure 4(b) depicts the band structure of ScFeO$_3$ within GGA+$U$ calculation and AFM structure. The band gap is about 2.41 eV with indirect nature. The band dispersion near Fermi energy is quite large for valence band and relatively small for conduction band. In order to see the Fe ion contribution more clearly, we have calculated the Fe atom projected band structure in Fig. 4(c). Figure 4(c) represent the down spin channel of band with contribution, of which the size of symbol is the amount of contribution of Fe atom in that band energy. There are two different Fe ions for AFM configuration, Fe1 represents up spin and Fe2 represents down spin. As can be clearly seen from figure (down spin channel) Fe2 is dominant for valence band (in other word, occupied band), whereas Fe1 is dominant for conduction band (unoccupied band). Up spin channel has exactly opposite contribution for Fe1 and Fe2. From this figure, it can be easily visualized that the spin configuration of Fe1 is fully up spin and that of Fe2 is fully down spin. Therefore, one could conclude that above mentioned Goodenough-Kanamori super-exchange type AFM is dominant for magnetic ground state configuration. The Atom projected density of state is plotted in Fig. 4(d). The Fe contribution is shown as red solid line for spin up and red dashed line for spin down channel. By integrating up to Fermi energy, the magnetic moments for Fe ion are close to 5.0 and -5.0 Bohr magneton, which are again consistent with Goodenough-Kanamori picture of $d^5$ configuration of Fe ion in ScFeO$_3$ system. It is interesting to note that the experimental result is consistent with our calculation [27, 28] within first order. However, in the experimental data, the small magnetic anisotropy is realized in the real system and the small degree of spin canting is observed, which we would like to answer in the following section.

E. Spin canting and magnetic anisotropy

In order to check spin canting and magnetic anisotropy, we have done spin non-collinear calculation [43]. Our calculation results are summarized in Fig. 5. First of all, we have checked the magnetic anisotropy as shown in Fig. 5(a). While keeping AFM structure optimized in section A, we have changed the easy axis along [111] and [1$\bar{1}$0] axis. In the inset of Fig. 5(a), the red arrow is spin direction of Fe1 ion and the blue arrow is that of Fe2 ion. For this calculation, we preserve the AFM configuration of two spins. We plot total energy difference as a function of spin angle $\theta$. Our calculation clearly indicate that the magnetic easy axis is along the [1$\bar{1}$0] plane as shown in Fig. 5(a). The energy is sinusoidal function and the energy difference is ~ 25.7 $\mu$eV/f. u. Then we have check the magnetic anisotropy along the six equivalent [1$\bar{1}$0] plane while maintaining AFM configuration as depicted in Fig. 5(b). As can be clearly seen, we could not found any magnetic anisotropy as a function of $\phi$ within the accuracy of our calculations.

Having the magnetic anisotropy results, we are now in a position to check any canting possibility as a result of spin-orbit interaction. We calculated the total energy difference as a function of canting angle $\delta$ as shown in Fig. 5(c). The canting angle $\delta/2$ is defined by the deviation of each spin angle from the plane formed by [01$\bar{1}$] axis and [1$\bar{1}$0] axis. When each spin canted the value of $\delta/2$, the total spin canting is $\delta$. If the canting angle becomes 180 $^{o}$, the system is regarded as spin configuration of FM. The energy scale for FM to AFM change is ~ 20 meV as discussed in previous section and that of magnetic anisotropy is order of ~ 0.025 meV (see Fig. 5(a) for example). Therefore, one needs very careful convergence check for such calculation. Our results for spin canting calculation are summarized in Fig. 5(c). The existence of symmetricity in the energy surface can be clearly seen and the canting angle can be estimated by fitting calculation data with polynomial function. The canting angle $\delta$ is estimated ~ 0.48 $^{o}$ and the energy lowering is order of 0.017 meV. Our finding of spin canting is consistent with experimental report of ref. 28. However, considering that the energy scale is about the limit of the accuracy of DFT calculations, further careful examination of spin canting with other experimental tools at low temperature is highly recommended.

IV. Conclusions

In conclusion, we have studied the phase transitions and physical properties of the $ScFeO_3$. We successfully explained the $a^-a^-a^-$, tilting pattern of *R3c* as ground state using generalized gradient approximation with on-site Column interaction (GGA+*U*: *U* ~ 4 eV) of Fe ion. With external pressure, the $a^-a^-a^-$ tilting pattern of *R3c* hexagonal structure transforms into the $a^-a^-c^+$ tilting of *Pnma* orthorhombic structure. The $a^-a^-a^-$ tilting pattern of *R3c* hexagonal structure is realized by two order parameters of $\Gamma_1^+$ mode and $\Gamma_2^-$ mode. The symmetry nature of structural phase transition is same as that observed in other perovskites such as $LiNbO_3$ and $LiOsO_3$. In the *R3c* hexagonal structure, $\Gamma_2^-$ mode is related with the existence of spontaneous polarization of 100.4 $\mu C/cm^2$. We also analyzed the AFM magnetic configurations are always lower energy state compared to ferromagnetic (FM) states. Our spin non-collinear calculation confirms the small amount of spin canting (~ 0.48 $^o$) further lower the ground state energy. The detailed of canting direction and their energetics are also analyzed. Our calculation demonstrates successful application of GGA+*U* method to the understanding of the physical properties of $ScFeO_3$ and the predictability of new material design.

**ACKNOWLEDGEMENTS**

This study was supported by the NSF of Korea (NSF-2015R1D1A1A01057589)

# References

[1] E. Bousquet, M. Dawber, M. Stucki, C. Lichtensteiger, P. Hermet, S. Gariglio, J. M. Triscone, and P. Ghosez, Nature **452**, 732 (2008).

[2] J. Suntivich, K. J. May, H. G. Gasteiger, J. B. Goodenough, and Y. Shao-Horn, Science **334**, 1383 (2011).

[3] Y. Kozuka, M. Kim, C. Bell, Bog G. Kim, Y. Hikita, and H. Y. Hwang, Nature **462**, 487 (2009).

[4] N. A. Benedek and C. J. Fennie, Phys. Rev. Lett. **106**, 107204 (2011).

[5] T. Fukushima, A. Stroppa, S. Picozzi, and J. M. Perez-Mato, Phys. Chem. Chem. Phys. **13**, 12186 (2011).

[6] X. Fan, W. Zheng, X. Chen, and D. J. Singh, Plos One **9**, e91423 (2014).

[7] A. Aoyama, K. Yamauchi, A. Iyama, S. Picozzi, K. Shimizu, and T. Kimura, Nature Comm. **5**, 4927 (2014).

[8] M. J. Pitcher, P. Mandal, M. S. Dyer, J. Alaria, P. Borisov, H. Niu, J. B. Claridge, and M. J. Rosseinsky, Science **347**, 420 (2015).

[9] H. Lee, S. W. Choeng, and Bog G. Kim, J. Solid State Chem. **228**, 214 (2015).

[10] J. M. Rondinelli and N. A. Spaldin, Adv. Mater. **23**, 3363 (2011)

[11] A. Malashevich and D. Vanderbilt, Phys. Rev. Lett. **101**, 037210 (2008).

[12] J. H. Lee and K. M. Rabe, Phs. Rev. Lett. **107**, 067601 (2011)

[13] S. Singh, J. T. Haraldsen, J. Xiong, E. M. Choi, P. Lu, D. Yi, X. D. Wen, J. Liu, H. Wang, Z. Bi, P. Yu, M. R. Fitzsimmons, J. L. MacManus-Driscoll, R. Ramesh, A. V. Balatsky, J. X. Zhu and Q. X. Jia, Phys. Rev. Lett. **113**, 047204 (2014)

[14] C. Bell, S. Harashima, Y. Kozuka, M. Kim, Bog G. Kim, Y. Hikita, and H. Y. Hwang, Phys. Rev. Lett. **103**, 226802 (2009).

[15] S. –W. Cheong and M. Mostovoy, Nature Mater. **6**, 13 (2007).

[16] T. Kimura, T. Goto, H. Shintani, K. Ishizaka, T. Arima, and Y. Tokura, Nature **426**, 55 (2003).

[17] M. Toyoda, K. Yamauchi, and T. Oguchi, Phys. Rev. B **87**, 224430 (2013).

[18] *Physics of ferroelectrics: a modern perspective*, edited by K. M. Rabe, C. H. Ahn, J. -M. Triscone, Top. Appl. Phys. Vol. 105 (Springer, Berlin 2007).

[19] R. D. Shannon, Acta Cryst. A **32,** 751 (1976).

[20] A. M. Glazer, Acta Cryst. B **28**, 3384 (1972).

[21] P. M. Woodward, Acta Cryst. B **53**, 44 (1997),

[21] C. J. Howard and H. T. Stokes, Acta Cryst. B **54**, 782 (1998)

[22] H. Sim, S. W. Cheong, and Bog G. Kim, Phys. Rev. B **88**, 014101 (2013),

[23] J. Young and J. M. Rondinelli, Chem. Mater. **25**, 4545 (2013).

[24] J. Hong and D. Vanderbilt, Phys. Rev. B **87**, 064104 (2013).

[25] L Bellaiche and J. Iniguez, Phys. Rev. B **88**, 014104 (2013).

[26] J. T. Schick, L. Jiang, D. Saldana-Greco, and A. M. Rappe, Phys. Rev. B **89**, 195304 (2014).

[27] M.-R. Li, U. Adem, S. R. C. McMitchell, Z. Xu, C. I. Thomas, J. E. Warren, D. V. Giap, H. Niu, X. Wan, R. G. Palgrave, F. Schiffmann, F. Cora, B. Slater, T. L. Burnett, M. G. Cain, A. M. Abakumov, G. van Tendeloo, M. F. Thomas, M. J. Rosseinsky, and J. B. Claridge, J. Am. Chem. Soc. **134**, 3737 (2012).

[28] T. Kawamoto, K. Fujita, I. Yamada, T. Matoba, S. J. Kim, P. Gao, X. Pan, S. D. Findlay, C. Tessel, H. Kageyama, A. J. Studer, J. Hester, T. Irifune, H. Akamatsu, and K. Tanaka, J. Am. Chem. Soc. **136**, 15291 (2014).

[29] J. B. Neatonm C. Ederer, U. V. Waghmare, N. A. Spaldin, and K. M. Rabe, Phys. Rev. B **71**, 014113 (2005).

[30] J. P. Perdew, K. Burke, and M. Ernzerhof, Phys. Rev. Lett. **77**, 3865 (1996); **78**, 1396 (E) (1997).

[31] S. L. Dudarev, G. A. Botton, S. Y. Savrasov, C. J. Humphreys and A. P. Sutton, Phys. Rev. B **57**, 1505 (1998).

[32] G. Kresse and J. Furthmuller, Phys. Rev. B **54**, 11169 (1996).

[33] G. Kresse and D. Joubert, Phys. Rev. B **59**, 1758 (1999).

[34] We have changed the U from 3-4 eV for consistent check and found the essential physics remain unchanged.

[35] H. J. Monkhorst and J. D. Pack, Phys. Rev. B **13**, 5188 (1976).

[36] ISOTROPY, http://stokes.byu.edu/isotropy.html.

[37] D. Orobengoa, C. Capillas, M. I. Aroyo, and J. M. Perez-Mato, J. Appl. Cryst. **42**, 820 (2009).

[38] R. D. King-Smith and D. Vanderbilt, Phys. Rev. B **47**, 1651(R) (1993); D. Vanderbilt and R. D. King-Smith, Phys. Rev. B **48**, 4442 (1993); R. Resta, Rev. Mod. Phys. **66**, 899 (1994).

[39] D. Hobbs, G. Kresse, and J. Hafner, Phys. Rev. B. **62**, 11556 (2000).

[40] K. Parinski, Z. Q. Li, and Y. Kawazoe, Phys. Rev. B **61**, 272 (2000).; Q. Peng and R. E. Cohen, Phys. Rev. B **83**, 220103 (2011).

[41] H. Sim and Bog G. Kim, Phys. Rev. B 89, 201107(R) (2014).

[42] J. B. Goodenough, J. Phys. Chem. Solids **6**, 287 (1958); J. Kanamori, J. Phys. Chem. Solids **10**, 87 (1959).

[43] D. Hobbs, G. Kresse, and J. Hafner, Phys. Rev. B **62**, 11556 (2000).

**Figure Captions**

Figure 1. Crystal structure and octahedral tilting of ScFeO$_3$ in (a) $R\bar{3}c$, (b) $R3c$, and (c) *Pnma* space group in [001] projection. Purple big sphere is Sc and red small sphere is Oxygen. (d) Sc and Fe ion displacement for *R3c* space group shown in primitive cell. (e) Calculated total energy vs. volume in formula unit scale with GGA+$U$ (Fe = 4 eV). Note that *R3c* has lowest energy state with largest volume at ground state.

Figure 2. (a) Pressure dependence of enthalpy of the orthorhombic *Pnma* Structure and the hexagonal *R3c* structure. (b) Volume as a function of applied pressure. (c) Pseudo-cubic lattice parameters vs applied pressure. (d) Tilting angle of $a^-$ and $c^+$ as a function of applied pressure. Note that the pressure induced phase transition can be clearly seen around 37.4 kbar.

Figure 3. (a) $\Gamma_1^+$ mode and (b) $\Gamma_2^-$ mode associated with phase transition from $R\bar{3}c$ to *R3c*. The crystal structure is shown in primitive cell and arrows in the figures represent atomic movement of symmetry modes. (c) The energy surface as a function of the mode amplitude ($\Gamma_1^+$ and $\Gamma_2^-$ mode) in ScFeO$_3$. Green squares are the total energy of the $\Gamma_1^+$ mode without a $\Gamma_2^-$ mode and the red squares are those for the $\Gamma_1^+$ mode in addition to the equilibrium displacement of the $\Gamma_2^-$ mode. (d) Polarization for adiabatic path from undistorted phase to distorted phase in AFM configuration. $P_S$ and $P_Q$ represent spontaneous polarization and polarization quanta.

Figure 4. (a) Energy vs. volume curve for two different magnetic configurations (FM vs. AFM). (b) Band structure of *R3c* AFM phase of ScFeO$_3$ within GGA+$U$ calculation. (c) Fe atom projected band structure of down spin channel for AFM configuration. Note that the sizes of symbol represent the contribution from Fe ion. (d) Atom projected partial density of states (PDOS) for *R3c* AFM structure. Signs in PDOS graph represent spin up (positive) and spin down (negative) channel.

Figure 5. Magnetic anisotropy energy curves of ScFeO$_3$ with respect to spin rotation (a) from the hard to easy axis and (b) within the easy axis plane. (c) Energy as a function of spin canting angle between two neighboring spins with opposite directions in the easy axis plane. The data are fitted by polynomial function (solid line).

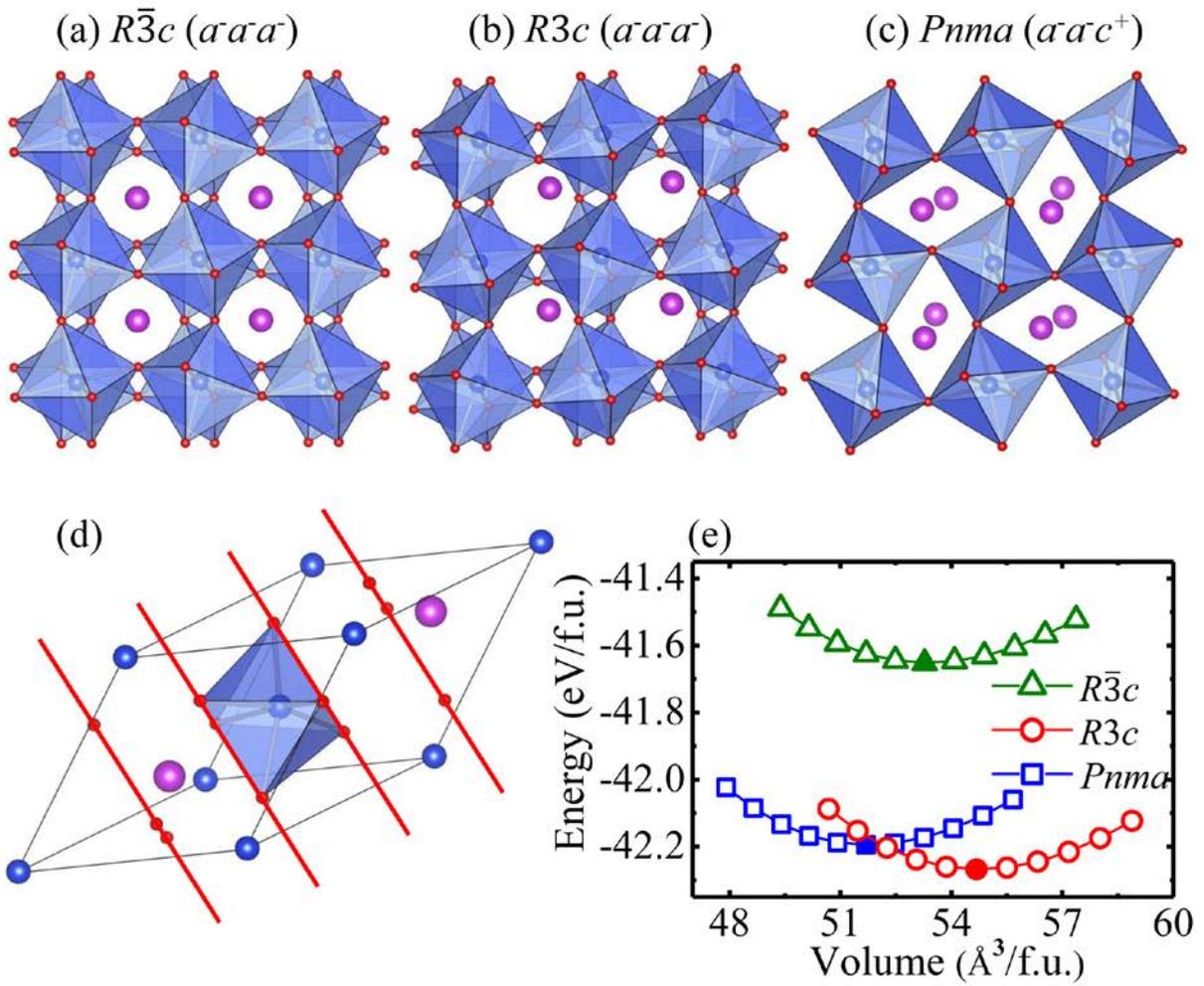

Figure 1 (Color online) Kim *et al.*

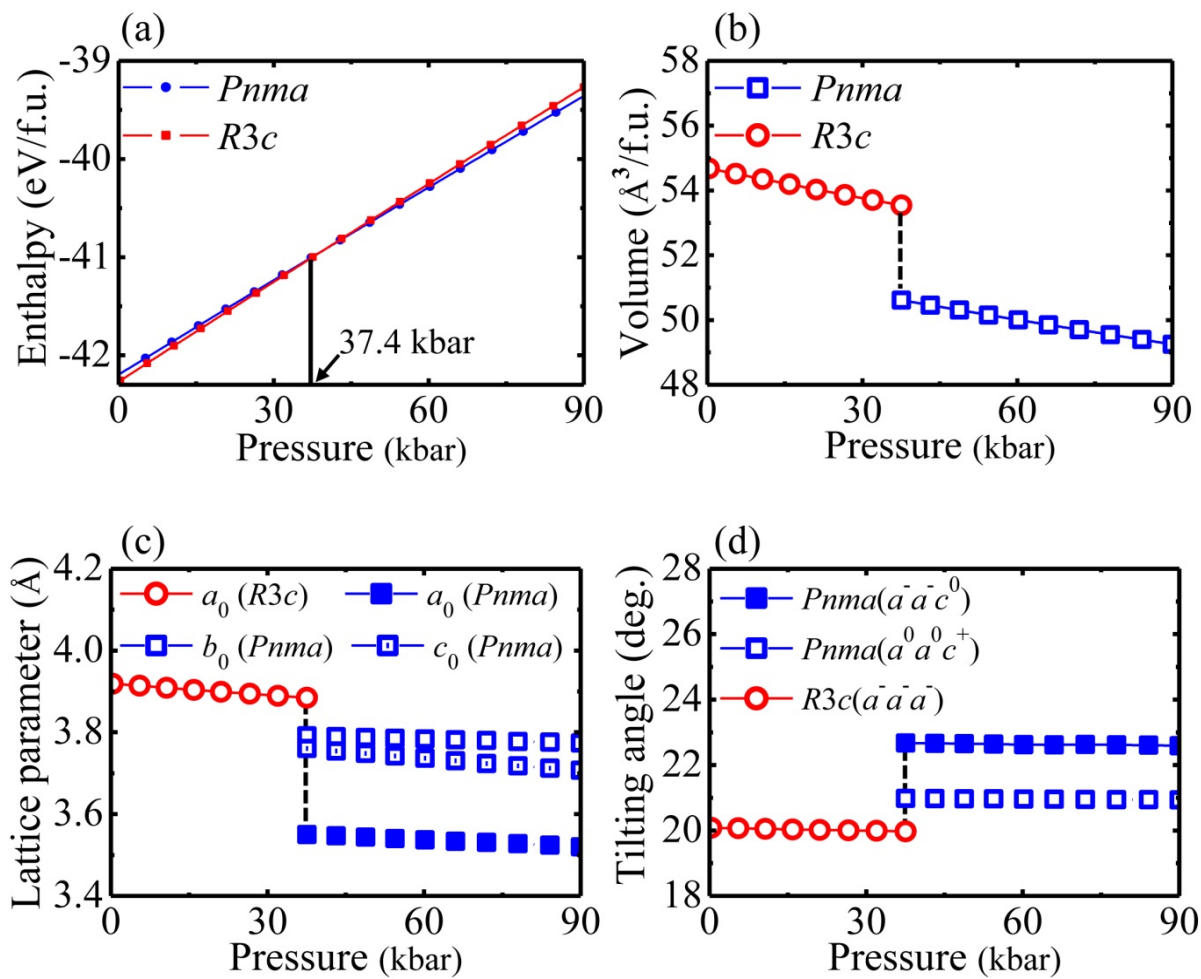

Figure 2 (Color online) Kim *et al.*

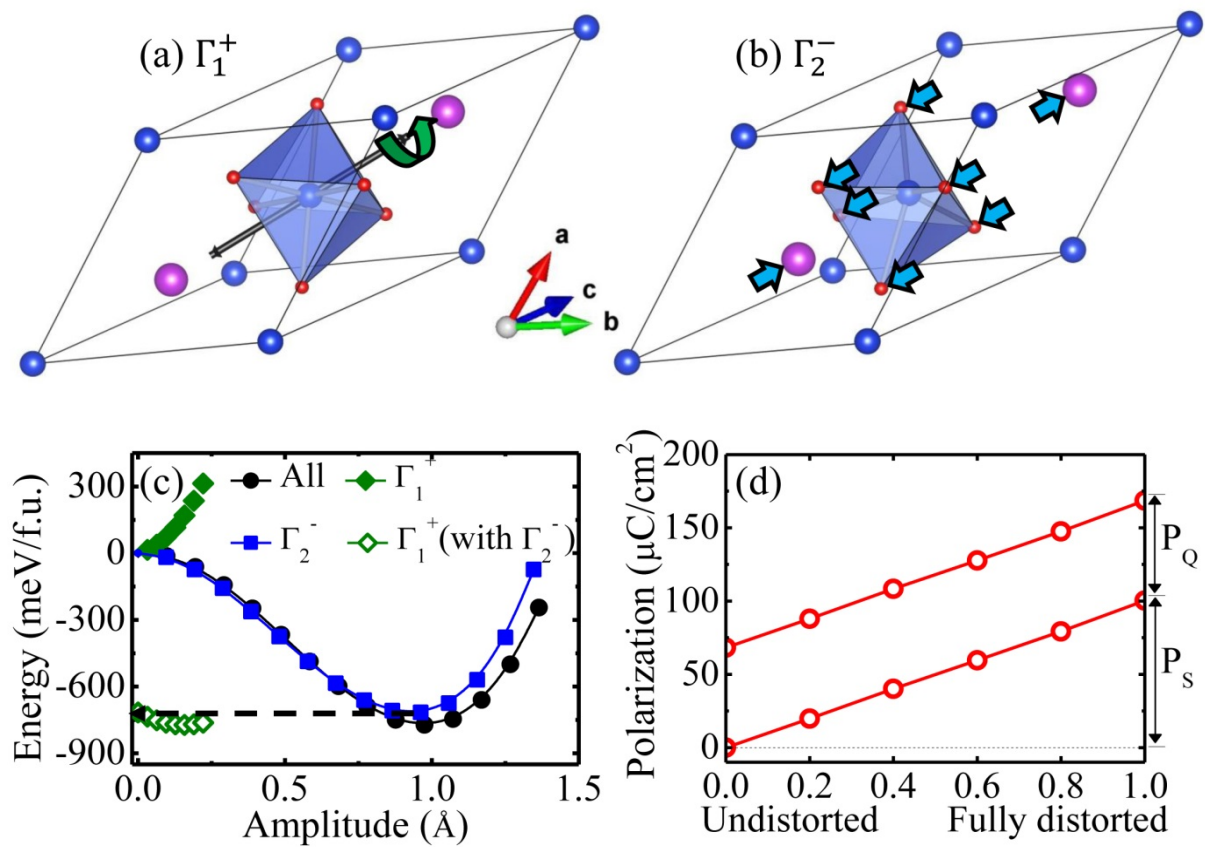

Figure 3 (Color online) Kim *et al.*

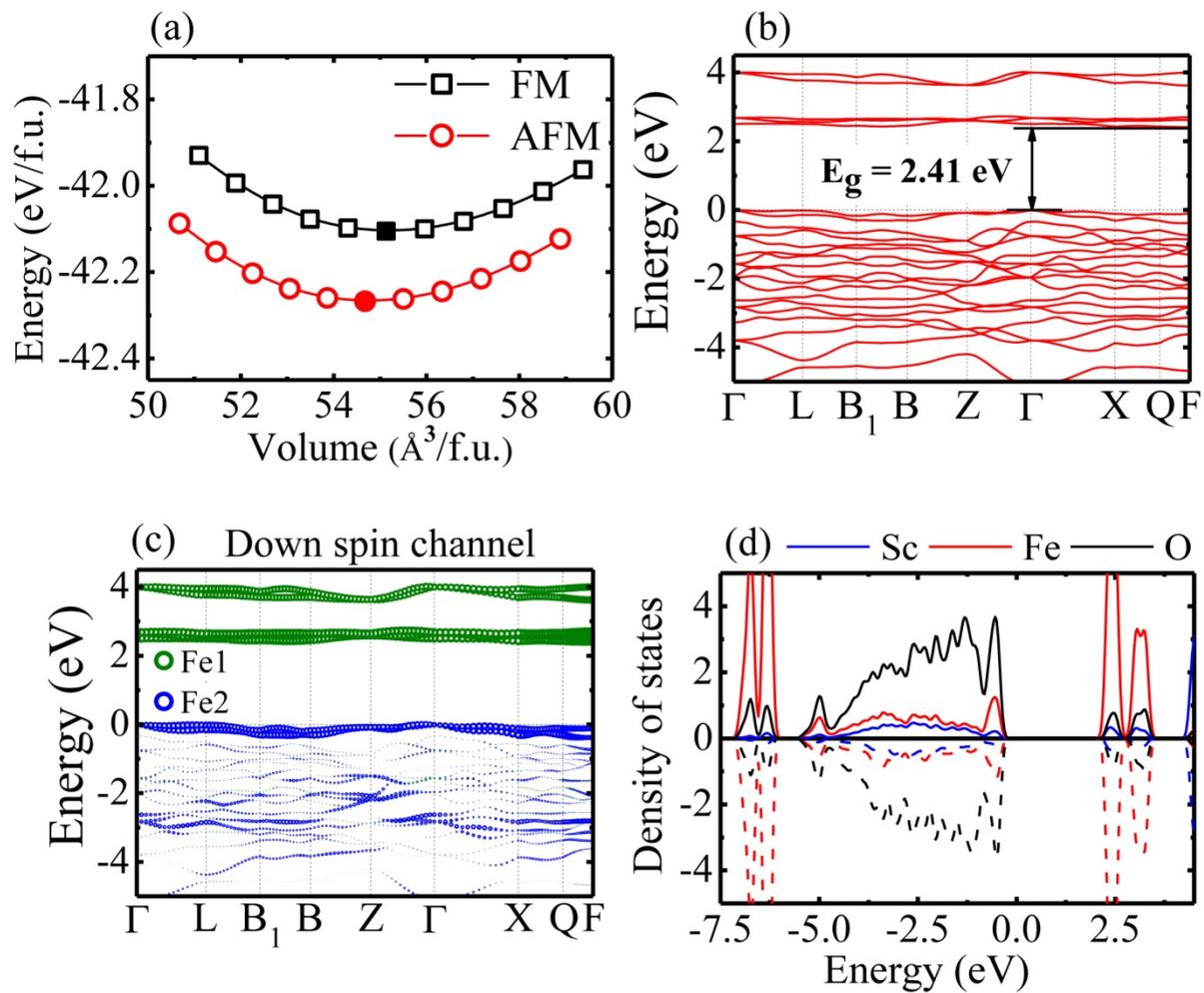

Figure 4 (Color online) Kim *et al.*

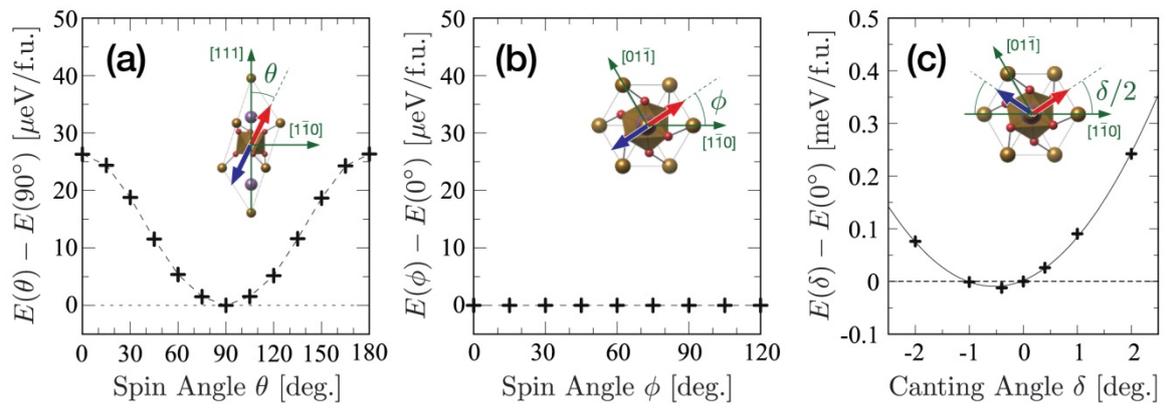

Figure 5 (Color online) Kim *et al.*

Table 1. Lattice constant and Wyckoff position of atoms of ScFeO$_3$ in $R\bar{3}c$, $R3c$, and $Pnma$ space group.

| Space group | | Wyckoff position | | | |
|---|---|---|---|---|---|
| $R\bar{3}c$ | Lattice constant (Å) | 5.3184 | 5.32184 | 13.0502 | |
| | Sc | 6a (0 0 1/4) | | | |
| | Fe | 6b (0 0 0) | | | |
| | O | 18e (x 0 1/4) | x = 0.3787 | | |
| $R3c$ | Lattice constant (Å) | 5.2085 | 5.2085 | 13.9646 | |
| | Sc | 6a (0 0 z) | z = 0.2879 | | |
| | Fe | 6a (0 0 z) | z = -0.0098 | | |
| | O | 18b (x y z) | x = 0.3770 | | |
| | | | y = 0.0207 | | |
| | | | z = 0.2297 | | |
| $Pnma$ | Lattice constant (Å) | 5.0489 | 5.3806 | 7.6074 | |
| | Sc | 4c (x y 1/4) | x = -0.0267 | | |
| | | | y = 0.0692 | | |
| | Fe | 4b (0 0 1/2) | | | |
| | O1 | 4c (x y 1/4) | x = 0.1430 | | |
| | | | y = 0.4372 | | |
| | O2 | 8d (x y z) | x = -0.1889 | | |
| | | | y = -0.1923 | | |
| | | | z = -0.4266 | | |

Table 1.  Kim *et al.*

Table 2. Amplitude, character, isotropy subgroup, and atomic displacement of $\Gamma_1^+$ and $\Gamma_2^-$ mode.

| Mode | Character | Isotropy subgroup | atom | Atomic displacement | | | Amplitude (Å) |
|---|---|---|---|---|---|---|---|
| | | | | δx | δy | δz | |
| $\Gamma_1^+$ | Displacement | $R\bar{3}c$ | Sc | 0.0000 | 0.0000 | 0.0000 | 0.1582 |
| | | | Fe | 0.0000 | 0.0000 | 0.0000 | |
| | | | O | 0.0768 | 0.0000 | 0.0000 | |
| $\Gamma_2^-$ | Rotation | $R3c$ | Sc | 0.0000 | 0.0000 | 0.0444 | 0.9620 |
| | | | Fe | 0.0000 | 0.0000 | 0.0040 | |
| | | | O | 0.0108 | 0.0216 | -0.0161 | |
| | | | | | Global distortion | | 0.9748 |

Table 2. Kim *et al.*